\documentclass[sigconf]{acmart}

\usepackage{booktabs} % For formal tables
\usepackage[ruled]{algorithm2e}
\usepackage{epsfig}
\usepackage{subfigure}

\newcommand{\calU}{\mathcal{U}}
\newcommand{\calI}{\mathcal{I}}
\newcommand{\GPR}{PR}
\newcommand{\bias}{B}
\newcommand{\disparity}{BD}
\newcommand{\userKNN}{\sc UserKNN}
\newcommand{\GULM}{\sc GULM}
\newcommand{\jsim}{\mathit{JSim}}

\renewcommand\footnotetextcopyrightpermission[1]{}
\setcopyright{none}
\begin{document}
\title{Bias Disparity in Recommendation Systems}

\author{Virginia Tsintzou}
\affiliation{
  \institution{Department of Computer}
  \institution{Science and Engineering}
  \institution{University of Ioannina}
}
\email{vtsintzou@cs.uoi.gr}

\author{Evaggelia Pitoura}
\affiliation{
  \institution{Department of Computer}
  \institution{Science and Engineering}
  \institution{University of Ioannina}
}
\email{pitoura@cs.uoi.gr}

\author{Panayiotis Tsaparas}
\affiliation{
  \institution{Department of Computer}
  \institution{Science and Engineering}
  \institution{University of Ioannina}
}
\email{tsap@cs.uoi.gr}

\begin{abstract}
Recommender systems have been applied successfully in a number of different domains, such as, entertainment, commerce, and employment. Their success lies in their ability to exploit the collective behavior of users in order to deliver highly targeted, personalized recommendations.
Given that recommenders learn from user preferences, they incorporate different \emph{biases}~\cite{bias-online} that users exhibit in the input data.
More importantly, there are cases where recommenders may amplify such biases, leading to the phenomenon of \emph{bias disparity}.
In this short paper, we present a preliminary experimental study on synthetic data, where we investigate different conditions under which a recommender exhibits bias disparity, and the long-term effect of recommendations on data bias.
We also consider a simple re-ranking algorithm for reducing bias disparity, and present some observations for data disparity on real data.

\end{abstract}

\maketitle

\section{Introduction}

Recommender systems have found applications in a wide range of domains, including e-commerce, entertainment, social media, news portals, and employment sites~\cite{CF1}.
One of the most popular classes of recommendation systems is collaborative filtering. Collaborative Filtering (CF) uses the collective behavior of all users over all items to infer the preferences of individual users for specific items~\cite{CF1}.
However, given the reliance of CF algorithms on the input preferences, they are susceptible to \emph{biases} that may appear in the input data.
In this work, we consider biases with respect to the preferences of specific groups of users (e.g., men and women) towards specific categories of items (e.g., different movie genres).

Bias in recommendations is not necessarily always problematic. For example, it is natural to expect gender bias when recommending clothes. However, gender bias is undesirable when recommending job postings, or information content. Furthermore, we want to avoid the case where the recommender system introduces bias in the data, by amplifying existing biases and reinforcing stereotypes. We refer to this phenomenon, where input and recommendation bias differ, as \emph{bias disparity}.

The problem of algorithmic bias, and its flip side, fairness in algorithms, has attracted considerable attention in the recent years~\cite{bias-tutorial,fairness-awareness}. Most existing work focuses on classification systems, while there is limited work on recommendation systems.
One type of recommendation bias that has been considered in the literature is popularity bias~\cite{celma-hits}. It has been observed that under some conditions popular items are more likely to be recommended leading to a rich get richer effect, and there are some attempts to correct this bias~\cite{DBLP:conf/recsys/KamishimaAAS14}.
Related to this is also the quest for diversity~\cite{diversity}, where the goal is to include different types of items in the recommendations.

These notions of fairness do not take into account the presence of different (protected) groups of users and different item categories that we consider in this work. In~\cite{burke:slim} they assume different groups of users and items, they define two types of bias and they propose a modification of the recommendation algorithm in~\cite{karypis} to ensure a fair output. Their work focuses on fairness, rather than bias disparity, and works with a specific algorithm. The notion of bias disparity is examined in~\cite{BCWS16} but in a classification setting.
Fairness in terms of correcting rating errors for specific groups of users was studied in \cite{nips17} for a matrix factorization CF recommender.

In this paper, we consider the problem of bias disparity in recommendation systems. More specifically:
\begin{itemize}
\item We define notions of bias and bias disparity for recommender systems.
\item Using synthetic data we study different conditions under which bias disparity may appear. We consider the effect of the iterative application of recommendation algorithms on the bias of the data.
\item We present some observations on bias disparity on real data, using the MovieLens\footnote{MovieLens 1M: https://grouplens.org/datasets/movielens/1m/} dataset.
\item We consider a simple re-ranking algorithm for correcting bias disparity and study it experimentally.
\end{itemize}

\section{Model}

\subsection{Definitions}
We consider a set of $n$ users $\calU$ and a set of $m$ items $\calI$.
We are given implicit feedback in a $n\times m$ matrix $S$, where $S(u,i) =1$ if user $u$ has selected item $i$, and zero otherwise. Selection may mean that user $u$ liked post $i$, or that $u$ purchased product $i$, or that $u$ watched video $i$.

We assume that users are associated with an attribute $A_U$, e.g., the gender of the user. The attribute $A_U$ partitions the users into \emph{groups}, that is, subsets of users with the same attribute value, e.g., men and women. We will typically assume that we have two groups and one of the groups is the \emph{protected group}. Similarly, we assume that items are associated with an attribute $A_I$, e.g., the genre of a movie, which partitions the items into \emph{categories}, that is, subsets of items with the same attribute value, e.g., action and romance movies.

Given the association matrix $S$, we define the input \emph{preference ratio} $\GPR_S(G,C)$ of group $G$ for category $C$ as
the fraction of selections from group $G$ that are in category $C$.
Formally:
\begin{equation}
\label{eqn:proportion}
\GPR_S(G,C) = \frac{\sum_{\substack{u\in G}}\sum_{\substack{i\in C}} S(u,i)}
{\sum_{\substack{u\in G}}\sum_{\substack{i\in \calI}} S(u,i)}
\end{equation}
This is essentially the conditional probability that a selection is in category $C$ given that it comes from a user in group $G$.

To assess the importance of this probability we compare it against the probability $P(C) = |C|/m$ of selecting from category $C$ when selecting uniformly at random. We define the \emph{bias} $\bias_S(G,C)$ of group $G$ for category $C$ as:
\begin{equation}
\label{eqn:bias}
\bias_S(G,C) = \frac{\GPR_S(G,C)}{P(C)}
\end{equation}
Bias values less than 1 denote \emph{negative bias}, that is, the group $G$ on average tends to select less often from category $C$, while bias values greater than 1 denote \emph{positive bias}, that is, that group $G$ favors category $C$ disproportionately to its size.

We assume that the recommendation algorithm outputs for each user $u$ a ranked list of $r$ items $R_u$. The collection of all recommendations can be represented as a binary matrix $R$, where $R(u,i) = 1$ if item $i$ is recommended for user $u$ and zero otherwise. Given matrix $R$, we can compute the output preference ratio of the recommendation algorithm, $\GPR_R(G,C)$,  of group $G$ for category $C$ using Eq. ~(\ref{eqn:proportion}), and the output bias $\bias_R(G,C)$ of group $G$ for category $C$.

To compare the bias of a group $G$ for a category $C$ in the input data $S$ and the recommendations $R$, we define the \emph{bias disparity}, that is, the relative change of the bias value.
\begin{equation}
\label{eqn:disparity}
\disparity(G,C) = \frac{\bias_R(G,C)-\bias_S(G,C)}{\bias_S(G,C)}
\end{equation}
%\end{definition}

Our definitions of preference ratios and bias are motivated by concepts of group proportionality, and group fairness considered in the literature~\cite{bias-tutorial,fairness-awareness}.

\subsection{The Recommendation Algorithm}

For the recommendations, in our experiments, we use a user-based $K$-Nearest-Neighbors ({\userKNN}) algorithm.
The {\userKNN} algorithm first  computes for each user, $u$, the set $N_K(u)$ of the $K$ most similar users to $u$.
For similarity, it uses the Jaccard similarity, $\jsim$, computed using the matrix $S$.
For user $u$ and item $i$ not selected by $u$, the algorithm computes a \emph{utility value}
\begin{equation}
V(u,i)=
\frac
{\sum_{\substack{n\in N_K(u)}} \jsim (u,n) S(n,i)}
{\sum_{\substack{n\in N_K(u)}}\jsim (u,n)}
\end{equation}
The utility value $V(u,i)$ is the fraction of the similarity scores of the top-$K$ most similar users to $u$ that have selected item $i$.
To recommend $r$ items to a user, the $r$ items with the highest utility values are selected.

\section{Bias Disparity on Synthetic Data}
In this section, we present experiments with synthetic data. Our goal is to study the conditions under which the {\userKNN} exhibits bias disparity.

\subsection{Synthetic data generation}
Users are split into two groups $G_1$ and $G_2$ of size $n_1$ and $n_2$ respectively, and items are partitioned into two categories $C_1$ and $C_2$ of size $m_1$ and $m_2$ respectively. We assume that users in $G_1$ tend to favor items in category $C_1$, while users in group $G_2$ tend to favor items in category $C_2$.
To quantify this preference, we give as input to the data generator two parameters $\rho_1,\rho_2$, where parameter $\rho_i$ determines the preference ratio $\GPR_S(G_i,C_i)$ of group $G_i$ for category $C_i$. For example, $\rho_1 = 0.7$ means that 70\% of the ratings of group $G_1$ are in category $C_1$.

The datasets we create consist of 1,000 users and 1,000 items.
We assume that  each user selects 5\% of the items in expectation and
we recommend $r = 10$ items per user.
The presented results are average values of 10 experiments.

We perform two different sets of experiments. In the first set, we examine the role of the preference ratios and in the second set the role of group and category sizes.

\subsection{Varying the preference ratios}
In these experiments, we create datasets with equal-size groups $G_1$ and $G_2$, and equal-size item categories $C_1$ and $C_2$, and we vary the preference ratios of the groups.

\vspace*{-0.1in}
\subsubsection{Symmetric Preferences:}
\label{sec:proportions}
\begin{figure*}[t]
\centering
\subfigure[$\GPR_R$, symmetric case]{
{\epsfig{file = 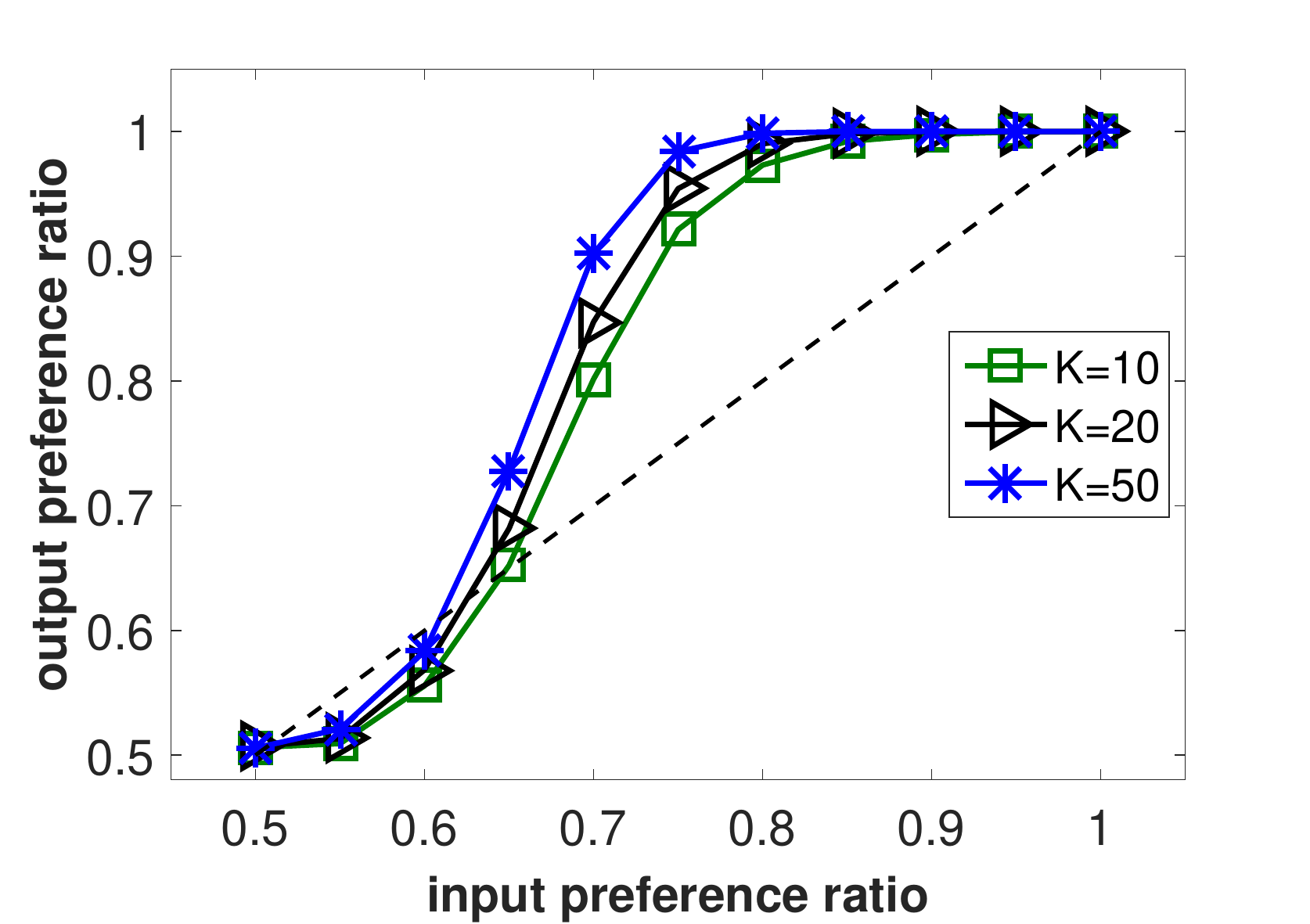, width = 0.2375\textwidth}}
}
\subfigure[Ratio of candidate items]{
{\epsfig{file = 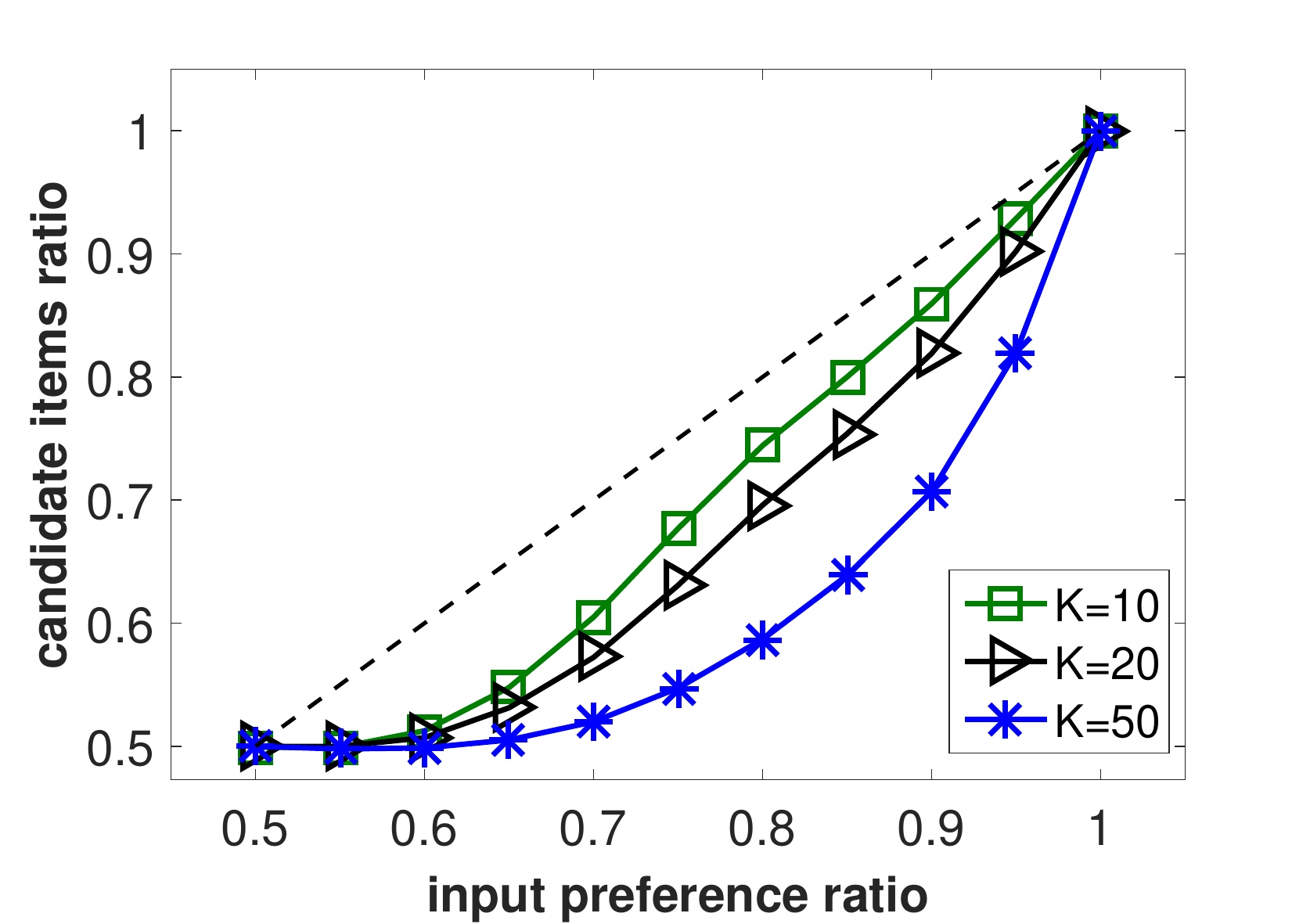, width = 0.2375\textwidth}}
}
\subfigure[$\GPR_R(G_1,C_1)$, asymmetric case]{
{\epsfig{file = 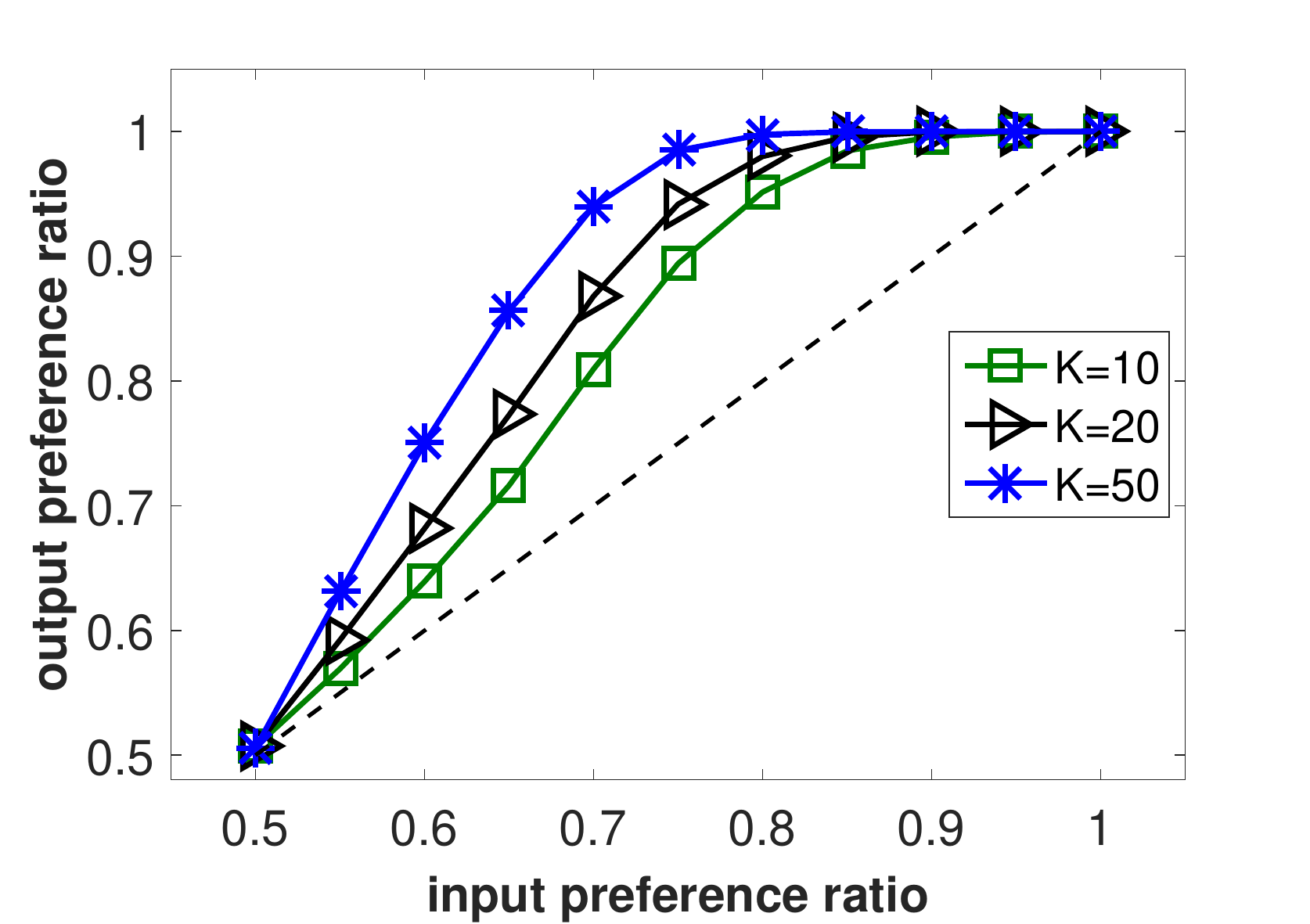, width = 0.2375\textwidth}}
}
\subfigure[$\GPR_R(G_2,C_2)$, asymmetric case]{
{\epsfig{file = 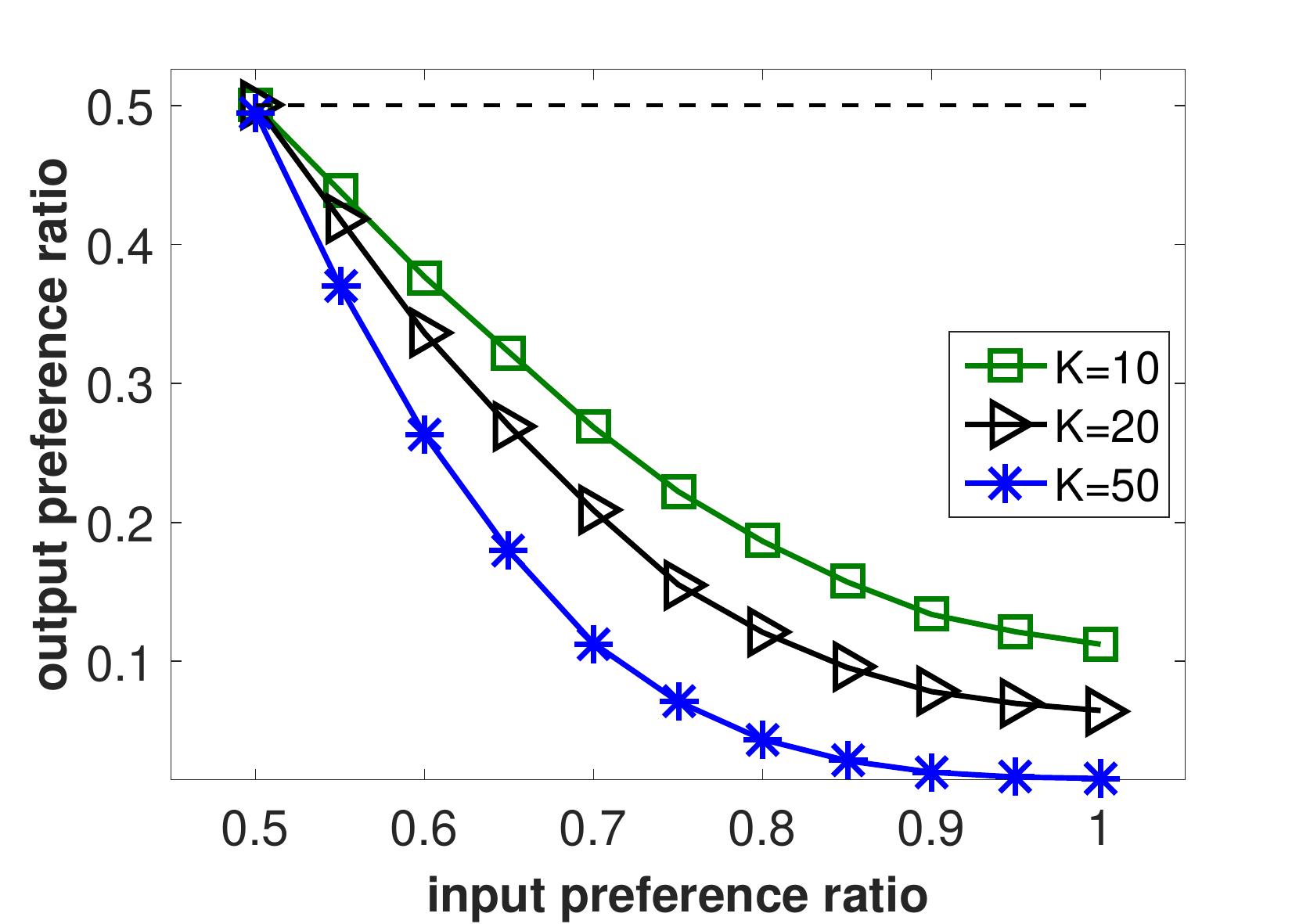, width = 0.2375\textwidth}}
}
\caption{Experiment with different preference ratios.}
\label{fig:preferences}
\end{figure*}

In the first experiment, we assume that the two groups $G_1$ and $G_2$ have the same preference ratios by setting $\rho_1=\rho_2=\rho$, where $\rho$ takes values from 0.5 to 1, in increments of 0.05.
In Figure~\ref{fig:preferences}(a), we plot the output preference ratio $\GPR_R(G_1, C_1)$ (eq. $\GPR_R(G_2, C_2)$) as a function of $\rho$. Note that in this experiment, bias is the preference ratio scaled by a factor of two. We report preference ratios to be more interpretable.
The dashed line shows when the output ratio is equal to the input ratio and thus there is no bias disparity. We consider different values for $K$, the number of neighbors.
A first observation is that when the input bias is small ($\GPR_S \leq 0.6$), the output bias decreases or stays the same. In this case, users have neighbors from both groups.
For higher input bias ($\GPR_S > 0.6$), we have a sharp increase of the output bias, which reaches its peak for $\GPR_S = 0.8$. In these cases, the recommender polarizes the two groups, recommending items only from their favored category.

In Figure~\ref{fig:preferences}(b), we report the preference ratio for all candidate items for recommendation for each user (i.e., all items having non zero utility).
Surprisingly, the candidate items  are less biased even for high values of the input bias. This shows that (a) utility proportional to user-similarity increases bias, (b) re-ranking may help in decreasing bias.

Increasing the value of K increases the output bias. Adding neighbors increases the strength of the signal, and the algorithm discriminates better between the items in the different categories. Understanding the role of $K$ is a subject for future study.

\setlength{\abovecaptionskip}{0pt}
\setlength{\belowcaptionskip}{-10pt}

\vspace*{-0.1in}
\subsubsection{Asymmetric Preferences:}
In this experiment, group $G_1$ has preference ratio $\rho_1$ ranging from 0.5 to 1 while $G_2$ has fixed preference ratio $\rho_2=0.5$, that is, $G_2$ is unbiased. In Figure~\ref{fig:preferences}, we show the recommendation preference ratio for groups $G_1$ (Figure ~\ref{fig:preferences}(c)) and $G_2$ (Figure~\ref{fig:preferences}(d)) as a function of $\rho_1$.

We observe that the output bias of group $G_1$ is amplified at a rate much higher than in Figure~\ref{fig:preferences}(a), while group $G_2$ becomes biased towards category $C_1$.
Surprisingly, the presence of the unbiased group $G_2$, rather than moderating the overall bias, it has an amplifying effect on the bias of $G_1$, more so than an opposite-biased group.
Furthermore, the unbiased group (Figure~\ref{fig:preferences}(d)) adopts the biases of the bias group.
This is due to the fact that the users in the unbiased group $G_2$ provide a stronger signal in favor of category $C_1$ compared to the symmetric case where group $G_2$ is biased over $C_2$. This reinforces the overall bias in favor of category $C_1$.

\subsection{Varying group and category sizes}

In this experiment we examine bias disparity  with unbalanced groups and categories.

\subsubsection{Varying Group Sizes:}
We first consider groups of uneven size. We set the size $n_1$ of $G_1$ to be a fraction $\phi$ of the number of all users $n$, ranging from 5\% to 95\%. Both groups have fixed preference ratio $\rho_1 = \rho_2 =0.7$.
Figure~\ref{fig:sizes}(a) shows the output recommendation preference ratio $\GPR_R(G_1,C_1)$ as a function of $\phi$. The plot of $\GPR_R(G_2,C_2)$ is the mirror image of this one, so we do not report it.

We observe that for $\phi \leq 0.3$ group $G_1$ has negative bias disparity ($\GPR_R(G_1,C_1) < 0.7$). That is, the small group is drawn by the larger group.
For medium values of $\phi$ in $[0.35,0.5]$ the bias of both groups is amplified, despite the fact that $G_1$ is smaller than $G_2$. The increase is larger for the larger group, but there is increase for the smaller group as well.

We also experimented with the case where $G_2$ is unbiased. In this case $G_2$ becomes biased towards $C_1$ even for $\phi = 0.05$, while the point at which the bias disparity for $G_1$ becomes positive is much earlier ($\phi \approx 0.2$). This indicates that a small biased group can have a stronger impact than a large unbiased one.

\subsubsection{Varying Category Sizes:}
\begin{figure}[ht]
\centering{
\subfigure[Group Size]
{\epsfig{file = 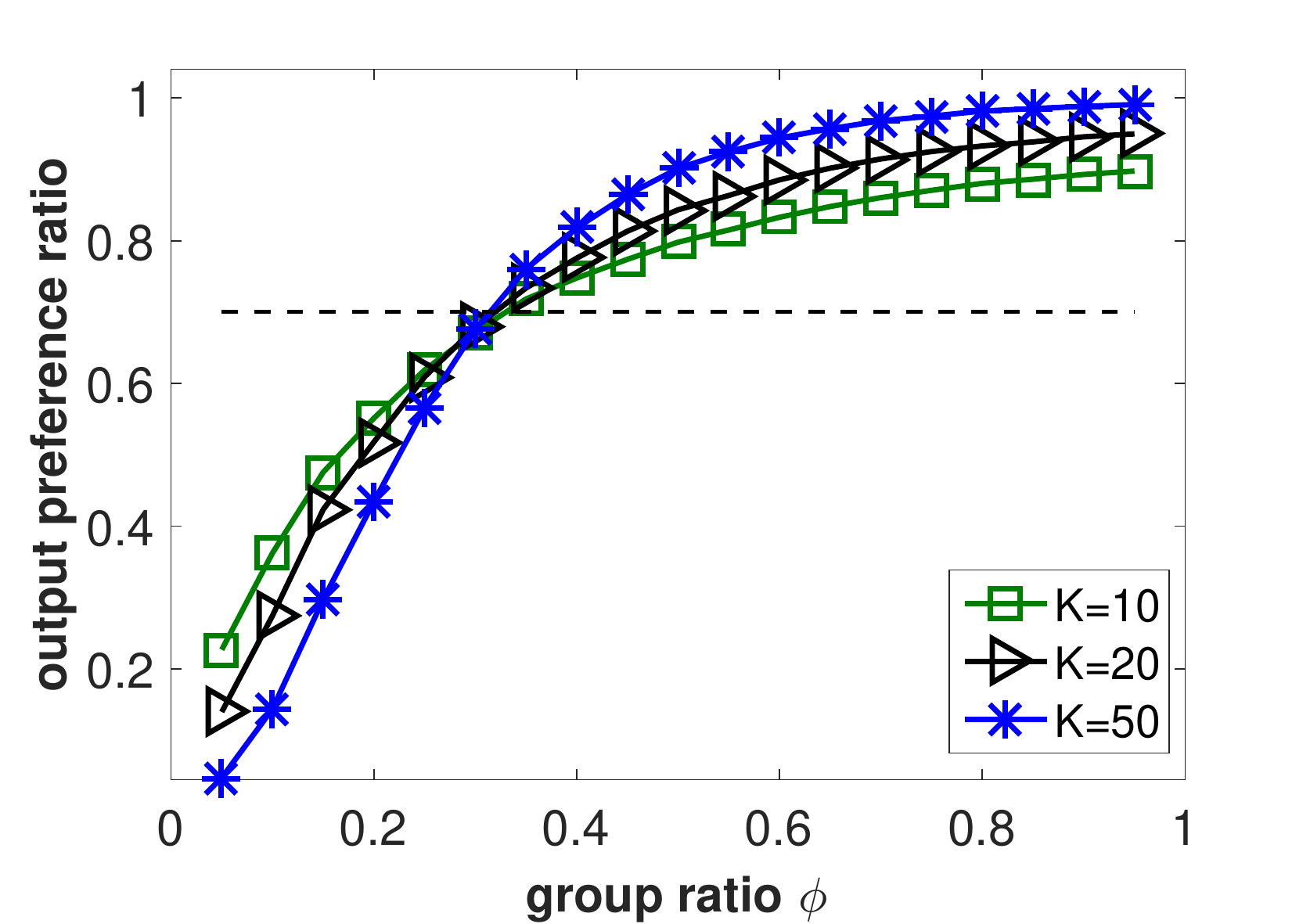, width = 0.47\columnwidth}
}
\subfigure[Category Size]
{\epsfig{file = 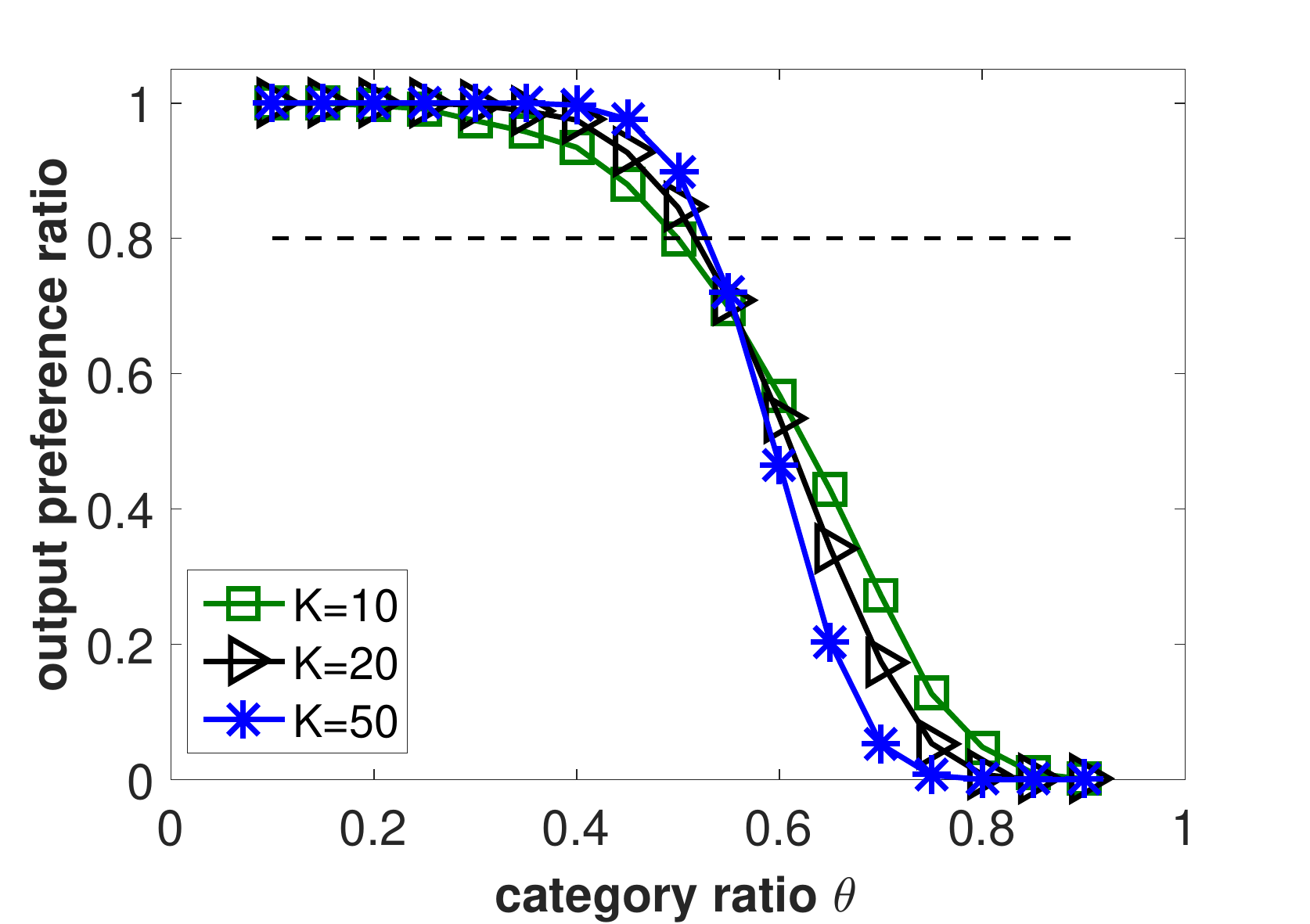, width = 0.47\columnwidth}
}
}
\caption{(a) Unbalanced group sizes, (b) Unbalanced category sizes; input preference ratio \boldmath{$\GPR_S(G_i,C_i) = 0.7$}.}
\label{fig:sizes}
\end{figure}

We now consider categories of uneven size. We set the size $m_1$ of $C_1$ to be a fraction $\theta$ of the number items $m$, ranging from 10\% to 90\%. We assume that both groups have fixed preference ratio $\rho_1 = \rho_2 =0.7$.
Figure~\ref{fig:sizes}(b) shows the recommendation preference ratio $\GPR_R(G_1,C_1)$ as a function of $\theta$. The plot of $\GPR_R(G_2,C_2)$ is again the mirror image of this one.

Note that as long as $\theta \leq 0.7$, group $G_1$ has positive bias (greater than 1) for category $C_1$ since bias is equal to $\rho_1/\theta$. However, it decreases as the size of the category increases. When the category size is not very large ($\theta \leq 0.5$), the output bias is amplified regardless of the category size. For $\theta > 0.7$, $G_1$ is actually biased in favor of $C_2$, and this is reflected in the output. There is an interesting range $[0.6,0.7]$ where $G_1$ is positively biased towards $C_1$ but its bias is weak, and thus the recommendation output is drawn to category $C_2$ by the more biased group.

\setlength{\tabcolsep}{1pt}

\subsection{Iterative Application of Recommendations}
\label{sec:iterative}

We observed bias disparity in the output of the recommendation algorithm. However, how does this affect the bias in the data? To study this we consider a scenario where the users accept (some of) the recommendations of the algorithm, and we study the long-term effect of the iterative application of the algorithm on the bias of the data. More precisely, at each iteration, we consider the top-$r$ recommendations of the algorithm ($r = 10$) to a user $u$, and we normalize their utility values, by the utility value of the top recommendation. We then assume that the user accepts a recommendation with probability equal to the normalized score. The accepted recommendations are added to the data, and they are fed as input to the next iteration of the recommendation algorithm.

\begin{figure}[ht]
\centering{
\setlength{\tabcolsep}{0pt}
\begin{tabular}{c c}
\multicolumn{2}{c}{
\subfigure{
\epsfig{file = 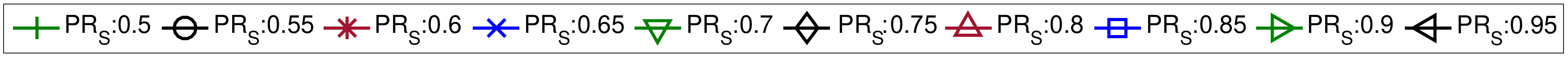, width = 0.9\columnwidth}
}} \\
\subfigure[UserKNN]{
\epsfig{file = 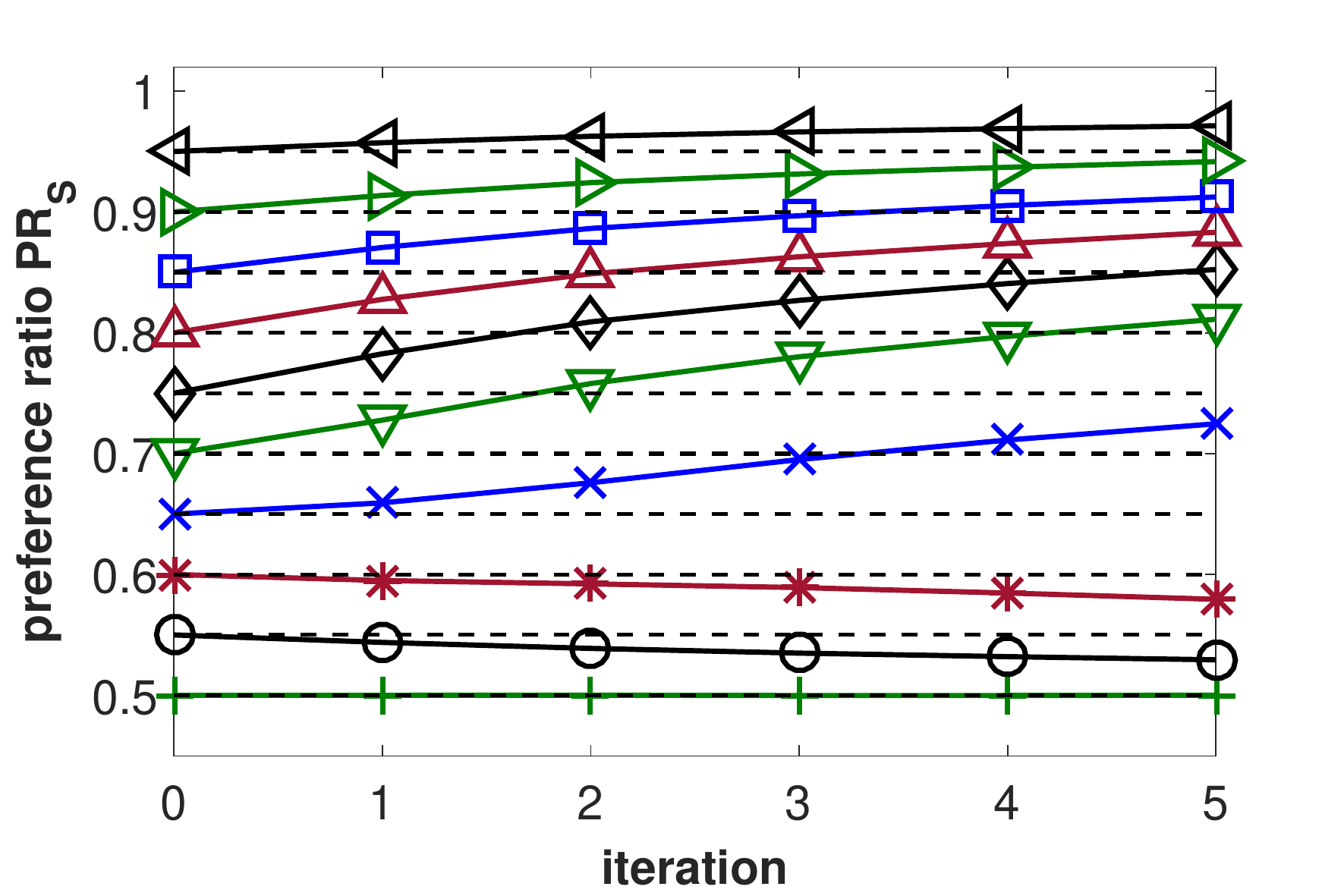, width = 0.475\columnwidth}
}
&
\subfigure[GULM]{
\epsfig{file = 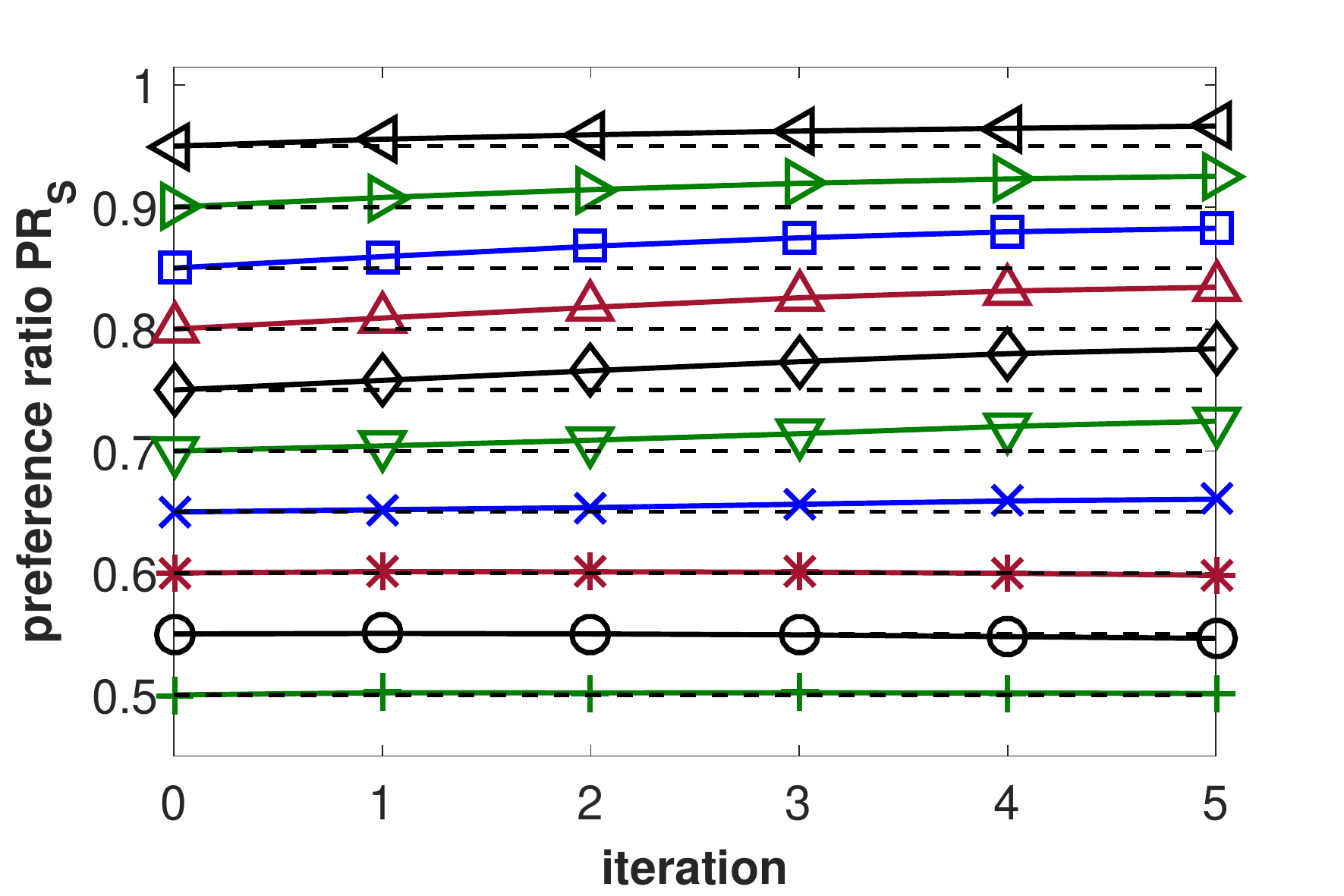, width = 0.475\columnwidth}
}
\end{tabular}
}
\caption{
The evolution of the preference ratio in the data for different input preference ratios ($\GPR_S$), after 5 iterations of (a) UserKNN and (b) GULM. Iteration 0 shows the original preference ratio of each experiment.
}
\label{fig:iterative}
\end{figure}

We apply this iterative algorithm on a dataset with two equally but oppositely biased groups, as described in Section~\ref{sec:proportions}. The results of this iterative experiment are shown in Figure~\ref{fig:iterative}(a), where we plot the average preference ratio for each iteration. Iteration 0 corresponds to the input data. In our experiment a user accepts on average 7 recommendations. For this experiment we set the number $K$ to 50.

We observe that even with the probabilistic acceptance of recommendations, there is a clear long-term effect of the recommendation bias. For small values of input bias, we observe a decrease, in line with the observations in Figure~\ref{fig:preferences}(a). For these values of bias, the recommender will result in reducing bias and smoothing out differences. The value of preference ratio 0.6 remains more or less constant, while for larger values the bias in the data increases. Therefore, for large values of bias the recommender has a reinforcing effect, which in the long term will lead to polarized groups of users.

\section{Bias disparity on Real Data}

In this experiment, we use the Movielens 1M dataset\footnote{MovieLens 1M: https://grouplens.org/datasets/movielens/1m/}.
We consider as categories the genres Action and Romance, with 468 and 463 movies. We extract a subset of users $\calU$ that have at least 90 ratings in these categories, resulting in 1,259 users.
Users in $\calU$ consist of 981 males and 278 females.

In Table~\ref{tab:gender}, we show the input/output bias and in parentheses the bias disparity for each group-category combination.
The right part of the table reports these numbers when the user groups are balanced, by selecting a random sample of 278 males.
We observe that males are biased in favor of Action movies while females prefer Romance movies. The application of {\userKNN} increases the output bias for males for which group the input bias is strong. Females are moderately biased in favor of Romance movies. Hence, their output bias is drawn to Action items.
We observe a very similar picture for balanced data, indicating that the changes in bias are not due to the group imbalance.

\setlength{\belowcaptionskip}{0pt}
\begin{table}[h]
	\caption{Gender bias on action and romance}
\begin{center}
\begin{small}
\begin{tabular}{ | c | c | c || c |c | }
 \hline
  & \multicolumn{2}{|c||}{Unbalanced Groups} & \multicolumn{2}{|c|}{Balanced Groups} \\
\hline
 & Action & Romance & Action & Romance \\
 \hline
M  & 1.39/1.67 (0.2) & 0.58/0.28 (-0.51) & 1.40/1.66 (0.18) & 0.57/0.29 (-0.49) \\
\hline
F  & 0.97/1.14 (0.17) & 1.03/0.85 (-0.17) & 0.97/1.08 (0.11) & 1.03/0.92 (-0.10) \\
 \hline
\end{tabular}
\end{small}
\label{tab:gender}
\end{center}
%\vspace{-15pt}
\end{table}

\section{Correcting Bias Disparity}

To address the problem of bias disparity, we consider an algorithm that performs post-processing of the recommendations.
Our goal is to adjust the set of items recommended to users so as to ensure that there is no bias disparity.
In addition, we would like the new recommendation set to have the maximum possible utility.

Abusing the notation, let $R$ denote the set of user-item pairs produced by our recommendation algorithm, where $(u,i) \in R$ denotes that $u$ was recommended item $i$.
We will refer to the pair $(u,i)$ as a recommendation. The set $R$ contains $r$ recommendations for each user, thus, $rn$ recommendations in total.
Let $V(R) = \sum_{(u,i) \in R} V(u,i)$ denote the total utility of the recommendations in set $R$. Since $R$ contains for each user $u$ the top-$r$ items with the highest utility, $R$ has the minimum utility loss.

We want to adjust the set $R$ so as to ensure that the bias of each group in $R$ is the same as the one in the input data. Since we have two categories, it suffices to have $\bias_R(G_i,C_i) = \bias_S(G_i,C_i)$.
Without loss of generality assume that $\bias_R(G_i,C_i) > \bias_S(G_i,C_i)$.
Let $\overline{C_i}$ denote the category other than $C_i$.

We decrease the output bias $\bias_R$ by swapping recommendations $(u,i)$ of category $C_i$ with recommendations $(u,j)$ of category $\overline{C_i}$.
We use a simple greedy algorithm that at each step swaps the pair of recommendations that incur the minimum utility loss. 
The utility loss incurred by swapping $(u,i)$ with $(u,j)$ is $V(u,i) - V(u,j)$.
The candidate swaps can be computed by pairing for each user $u$ the lowest-ranked recommendation $(u,i)$ in $R$ from category $C_i$, with the highest ranked recommendation $(u,j)$ not in $R$ from category $\overline{C_i}$.
We perform swaps like that until the desired number of swaps has been performed.
This algorithm is efficient, and it is easy to show that it is optimal, in the sense that it will produce the set of recommendations with the highest utility among all sets with
no bias disparity. We  refer to this algorithm as the {\GULM} (Group Utility Loss Minimization) algorithm.

By design, when we apply the {\GULM} algorithm on the output of the recommendation algorithm, we eliminate bias disparity (modulo rounding errors) in the recommendations. We consider the iterative application of the recommendation algorithm, in the setting described in Section~\ref{sec:iterative}, again
assuming that the probability of a recommendation being accepted depends on its utility.
The results are shown in Figure~\ref{fig:iterative}(b). For values of preference ratio up to 0.65, we observe that bias remains more or less constant after re-ranking. For larger values, there is some noticeable increase in the bias, albeit significantly smaller than before re-ranking. The increase is due to the fact that the recommendations introduced by {\GULM} have low probability to be accepted.

\section{Conclusions}

In this short paper, we performed a preliminary study of bias disparity in recommender systems,
and the conditions under which it may appear.
We view this analysis as a first step towards a systematic analysis of the factors that cause bias disparity. We intend to investigate more recommendation algorithms, and the case of numerical, rather than unary, ratings. We also want to better understand how the conditions we studied appear in real data.

\bibliographystyle{ACM-Reference-Format}
\bibliography{biasShortPaper-bib}


\begin{thebibliography}{11}

%%% the Web address.

\ifx \showCODEN    \undefined \def \showCODEN     #1{\unskip}     \fi
\ifx \showDOI      \undefined \def \showDOI       #1{#1}\fi
\ifx \showISBNx    \undefined \def \showISBNx     #1{\unskip}     \fi
\ifx \showISBNxiii \undefined \def \showISBNxiii  #1{\unskip}     \fi
\ifx \showISSN     \undefined \def \showISSN      #1{\unskip}     \fi
\ifx \showLCCN     \undefined \def \showLCCN      #1{\unskip}     \fi
\ifx \shownote     \undefined \def \shownote      #1{#1}          \fi
\ifx \showarticletitle \undefined \def \showarticletitle #1{#1}   \fi
\ifx \showURL      \undefined \def \showURL       {\relax}        \fi
% invisible to TeX
\providecommand\bibfield[2]{#2}
\providecommand\bibinfo[2]{#2}
\providecommand\natexlab[1]{#1}
\providecommand\showeprint[2][]{arXiv:#2}

\bibitem[\protect\citeauthoryear{Burke, Sonboli, and Ordonez-Gauger}{Burke
  et~al\mbox{.}}{2018}]%
        {burke:slim}
\bibfield{author}{\bibinfo{person}{Robin Burke}, \bibinfo{person}{Nasim
  Sonboli}, {and} \bibinfo{person}{Aldo Ordonez-Gauger}.}
  \bibinfo{year}{2018}\natexlab{}.
\newblock \showarticletitle{Balanced Neighborhoods for Multi-sided Fairness in
  Recommendation}. In \bibinfo{booktitle}{\emph{Proceedings of the 1st
  Conference on Fairness, Accountability and Transparency}}
  \emph{(\bibinfo{series}{Proceedings of Machine Learning Research})},
  \bibfield{editor}{\bibinfo{person}{Sorelle~A. Friedler} {and}
  \bibinfo{person}{Christo Wilson}} (Eds.), Vol.~\bibinfo{volume}{81}.
  \bibinfo{publisher}{PMLR}.
\newblock


\bibitem[\protect\citeauthoryear{Celma and Cano}{Celma and Cano}{2008}]%
        {celma-hits}
\bibfield{author}{\bibinfo{person}{\`{O}scar Celma} {and}
  \bibinfo{person}{Pedro Cano}.} \bibinfo{year}{2008}\natexlab{}.
\newblock \showarticletitle{From Hits to Niches?: Or How Popular Artists Can
  Bias Music Recommendation and Discovery}. In
  \bibinfo{booktitle}{\emph{Proceedings of the 2Nd KDD Workshop on Large-Scale
  Recommender Systems and the Netflix Prize Competition}}
  \emph{(\bibinfo{series}{NETFLIX '08})}. \bibinfo{publisher}{ACM},
  \bibinfo{pages}{5:1--5:8}.
\newblock


\bibitem[\protect\citeauthoryear{Dwork, Hardt, Pitassi, Reingold, and
  Zemel}{Dwork et~al\mbox{.}}{2012}]%
        {fairness-awareness}
\bibfield{author}{\bibinfo{person}{Cynthia Dwork}, \bibinfo{person}{Moritz
  Hardt}, \bibinfo{person}{Toniann Pitassi}, \bibinfo{person}{Omer Reingold},
  {and} \bibinfo{person}{Richard Zemel}.} \bibinfo{year}{2012}\natexlab{}.
\newblock \showarticletitle{Fairness Through Awareness}. In
  \bibinfo{booktitle}{\emph{Proceedings of the 3rd Innovations in Theoretical
  Computer Science Conference}} \emph{(\bibinfo{series}{ITCS '12})}.
  \bibinfo{publisher}{ACM}, \bibinfo{pages}{214--226}.
\newblock
\showISBNx{978-1-4503-1115-1}


\bibitem[\protect\citeauthoryear{Hajian, Bonchi, and Castillo}{Hajian
  et~al\mbox{.}}{2016}]%
        {bias-tutorial}
\bibfield{author}{\bibinfo{person}{Sara Hajian}, \bibinfo{person}{Francesco
  Bonchi}, {and} \bibinfo{person}{Carlos Castillo}.}
  \bibinfo{year}{2016}\natexlab{}.
\newblock \showarticletitle{Algorithmic Bias: From Discrimination Discovery to
  Fairness-aware Data Mining}. In \bibinfo{booktitle}{\emph{Proceedings of the
  22Nd ACM SIGKDD International Conference on Knowledge Discovery and Data
  Mining}} \emph{(\bibinfo{series}{KDD '16})}. \bibinfo{publisher}{ACM},
  \bibinfo{pages}{2125--2126}.
\newblock
\showISBNx{978-1-4503-4232-2}


\bibitem[\protect\citeauthoryear{Kamishima, Akaho, Asoh, and Sakuma}{Kamishima
  et~al\mbox{.}}{2014}]%
        {DBLP:conf/recsys/KamishimaAAS14}
\bibfield{author}{\bibinfo{person}{Toshihiro Kamishima},
  \bibinfo{person}{Shotaro Akaho}, \bibinfo{person}{Hideki Asoh}, {and}
  \bibinfo{person}{Jun Sakuma}.} \bibinfo{year}{2014}\natexlab{}.
\newblock \showarticletitle{Correcting Popularity Bias by Enhancing
  Recommendation Neutrality}. In \bibinfo{booktitle}{\emph{Poster Proceedings
  of the 8th {ACM} Conference on Recommender Systems, RecSys 2014, Foster City,
  Silicon Valley, CA, USA, October 6-10, 2014}}.
\newblock


\bibitem[\protect\citeauthoryear{Kunaver and Porl}{Kunaver and Porl}{2017}]%
        {diversity}
\bibfield{author}{\bibinfo{person}{Matev Kunaver} {and} \bibinfo{person}{Toma
  Porl}.} \bibinfo{year}{2017}\natexlab{}.
\newblock \showarticletitle{Diversity in Recommender Systems A Survey}.
\newblock \bibinfo{journal}{\emph{Know.-Based Syst.}} \bibinfo{volume}{123},
  \bibinfo{number}{C} (\bibinfo{date}{May} \bibinfo{year}{2017}),
  \bibinfo{pages}{154--162}.
\newblock
\showISSN{0950-7051}


\bibitem[\protect\citeauthoryear{Ning and Karypis}{Ning and Karypis}{2011}]%
        {karypis}
\bibfield{author}{\bibinfo{person}{Xia Ning} {and} \bibinfo{person}{George
  Karypis}.} \bibinfo{year}{2011}\natexlab{}.
\newblock \showarticletitle{SLIM: Sparse Linear Methods for Top-N Recommender
  Systems}. In \bibinfo{booktitle}{\emph{Proceedings of the 2011 IEEE 11th
  International Conference on Data Mining}} \emph{(\bibinfo{series}{ICDM
  '11})}. \bibinfo{publisher}{IEEE Computer Society},
  \bibinfo{pages}{497--506}.
\newblock


\bibitem[\protect\citeauthoryear{Pitoura, Tsaparas, Flouris, Fundulaki,
  Papadakos, Abiteboul, and Weikum}{Pitoura et~al\mbox{.}}{2017}]%
        {bias-online}
\bibfield{author}{\bibinfo{person}{Evaggelia Pitoura},
  \bibinfo{person}{Panayiotis Tsaparas}, \bibinfo{person}{Giorgos Flouris},
  \bibinfo{person}{Irini Fundulaki}, \bibinfo{person}{Panagiotis Papadakos},
  \bibinfo{person}{Serge Abiteboul}, {and} \bibinfo{person}{Gerhard Weikum}.}
  \bibinfo{year}{2017}\natexlab{}.
\newblock \showarticletitle{On Measuring Bias in Online Information}.
\newblock \bibinfo{journal}{\emph{CoRR}}  \bibinfo{volume}{abs/1704.05730}
  (\bibinfo{year}{2017}).
\newblock


\bibitem[\protect\citeauthoryear{Su and Khoshgoftaar}{Su and
  Khoshgoftaar}{2009}]%
        {CF1}
\bibfield{author}{\bibinfo{person}{Xiaoyuan Su} {and} \bibinfo{person}{Taghi~M.
  Khoshgoftaar}.} \bibinfo{year}{2009}\natexlab{}.
\newblock \showarticletitle{A Survey of Collaborative Filtering Techniques}.
\newblock \bibinfo{journal}{\emph{Adv. in Artif. Intell.}}
  \bibinfo{volume}{2009} (\bibinfo{date}{Jan.} \bibinfo{year}{2009}).
\newblock
\showISSN{1687-7470}


\bibitem[\protect\citeauthoryear{Yao and Huang}{Yao and Huang}{2017}]%
        {nips17}
\bibfield{author}{\bibinfo{person}{Sirui Yao} {and} \bibinfo{person}{Bert
  Huang}.} \bibinfo{year}{2017}\natexlab{}.
\newblock \showarticletitle{Beyond Parity: Fairness Objectives for
  Collaborative Filtering}.
\newblock \bibinfo{journal}{\emph{CoRR}}  \bibinfo{volume}{abs/1705.08804}
  (\bibinfo{year}{2017}).
\newblock


\bibitem[\protect\citeauthoryear{Zhao, Wang, Yatskar, Ordonez, and Chang}{Zhao
  et~al\mbox{.}}{2017}]%
        {BCWS16}
\bibfield{author}{\bibinfo{person}{Jieyu Zhao}, \bibinfo{person}{Tianlu Wang},
  \bibinfo{person}{Mark Yatskar}, \bibinfo{person}{Vicente Ordonez}, {and}
  \bibinfo{person}{Kai{-}Wei Chang}.} \bibinfo{year}{2017}\natexlab{}.
\newblock \showarticletitle{Men Also Like Shopping: Reducing Gender Bias
  Amplification using Corpus-level Constraints}.
\newblock \bibinfo{journal}{\emph{CoRR}}  \bibinfo{volume}{abs/1707.09457}
  (\bibinfo{year}{2017}).
\newblock


\end{thebibliography}

\end{document}